\documentclass[epj]{webofc}
\usepackage[utf8]{inputenc}
\usepackage[varg]{txfonts}   % Web of Conferences font
\usepackage{booktabs}
\usepackage{xcolor}
\definecolor{darkred}{rgb}{0.4,0.0,0.0}
\definecolor{darkgreen}{rgb}{0.0,0.4,0.0}
\definecolor{darkblue}{rgb}{0.0,0.0,0.4}
\usepackage[bookmarks,linktocpage,colorlinks,
    linkcolor = darkred,
    urlcolor  = darkblue,
    citecolor = darkgreen]{hyperref}
\usepackage{subfigure}
\wocname{EPJ Web of Conferences}
\woctitle{Lattice2017}
\def\mev{\mathrm{MeV}}

\begin{document}
%%%%%%%%%%%%%%%%%%%%%%%%%%%%%%%%%%%%%%%%%%%%%%%%%%%%%%%%%%%%%%%%%%%%%%%%%%%%%
%
\selectlanguage{english}
%----------------------------------------------------------------------------
\title{%
Glueball relevant study on isoscalars from $N_f=2$ lattice QCD
}
%----------------------------------------------------------------------------
\author{%
\firstname{Wei} \lastname{Sun}\inst{1,2}\fnsep\thanks{Speaker, \email{sunw@ihep.ac.cn}} \and
\firstname{Long-Cheng} \lastname{Gui}\inst{3,4} \and
\firstname{Ying}  \lastname{Chen}\inst{1,2} \and
\firstname{Ming} \lastname{Gong}\inst{1} \and
%\firstname{Chuan} \lastname{Liu}\inst{5,6} \and
%\firstname{Yu-Bin} \lastname{Liu}\inst{7} \and
\firstname{Zhaofeng} \lastname{Liu}\inst{1,2}
%\firstname{Jian-Ping} \lastname{Ma}\inst{8} \and
%\firstname{Jian-Bo} \lastname{Zhang}\inst{9} \\
%for the CLQCD Collaboration
% etc.
}
%----------------------------------------------------------------------------
\institute{%
Institute of High Energy Physics, Chinese Academy of Sciences, Beijing 100049, P.R. China
\and
School of Physics, University of Chinese Academy of Sciences, Beijing 100049, P.R. China
\and
Department of Physics and Synergetic Innovation Center for Quantum Effects and Applications, Hunan Normal University, Changsha 410081, P.R. China
\and
Key Laboratory of Low-Dimensional Quantum Structures and Quantum Control of Ministry of Education, Changsha 410081, P.R. China
%\and
%School of Physics and Center for High Energy Physics, Peking University, Beijing 100871, P.R. China
%\and
%School of Physics, Nankai University, Tianjin 300071, P.R. China
%\and
%Institute of Theoretical Physics, Chinese Academy of Sciences, Beijing 100080, P.R. China
%\and
%Department of Physics, Zhejiang University, Hangzhou, Zhejiang 310027, P.R. China
}
%----------------------------------------------------------------------------
\abstract{%
 We perform a glueball-relevant study on isoscalars based on  anisotropic $N_f=2$ lattice QCD gauge configurations. In the scalar channel, we identify the ground state obtained through gluonic operators to be a
 single-particle state through its dispersion relation. When $q\bar{q}$ operator is included, we find the mass of this state does not change,
 and the $q\bar{q}$ operator couples very weakly to this state. So this state is most likely a glueball state.
  For pseudoscalars, along with the exiting lattice results, our study implies that both the conventional $q\bar{q}$ state $\eta_2$ (or $\eta'$ in flavor
  $SU(3)$) and a heavier glueball-like state with a mass of roughly 2.6 GeV exist in the spectrum of lattice QCD with dynamical quarks.
}
%----------------------------------------------------------------------------
\maketitle
%----------------------------------------------------------------------------
\section{Introduction}\label{intro}

The Quantum Chromodynamics (QCD) predicts the existence of glueballs, which are exotic hadrons made up of gluons. Glueballs are well-defined objects
in the pure gauge theory or QCD in the quenched approximation (QQCD). The glueball spectrum has been extensively investigated by the QQCD lattice studies which
predicts the masses of lowest lying glueballs, such as the scalar, tensor, and pseudoscalar glueballs, are ranging from 1.5 GeV to 3.0 GeV~\cite{Morningstar:1999rf,Chen:2005mg}. Intuitively, for the lattice QCD simulations with dynamical quarks, the definition of glueballs is ambiguous since
glueballs can mix with conventional $q\bar{q}$ mesons and even multi-hadron states when the vacuum polarization of sea quarks is switched on. However, several
exploratory lattice studies with dynamical quarks observe that the mass spectrum of the lowest-lying glueballs is compatible with the results in the quenched
approximation and shows small unquenched effects if only the gluonic interpolation operators are adopted in the calculation ~\cite{Bali:2000vr,Hart:2001fp,Richards:2010ck,Gregory:2012hu}.

Recently, we also performed a similar study on glueball spectrum in $N_f=2$ QCD on two anisotropic lattices~\cite{Sun:2017ipk} with the pion mass 650 MeV and 938 MeV, respectively. In the tensor channel, we obtain the mass of ground state to be roughly 2.4 GeV,
which is compatible with the tensor glueball mass from QQCD studies. For the pseudoscalar, by using the conventional gluonic operators, the
ground state mass is determined to be roughly 2.6 GeV, which is also consistent with the pseudoscalar glueball mass derived in QQCD. However, if we use
the topological charge density (defined through Wilson loops) as the lattice interpolation operator for the pseudoscalar, we get a state with mass
roughly 0.89 GeV and 1.2 GeV  for these two pion masses respectively, which may correspond to the $\eta_2$ meson. The result in scalar channel
is somewhat ambiguous. The correlator of gluonic operators gives the ground state mass to be 1.4 GeV or so, which is smaller than the scalar glueball
mass from QQCD, but is close to the $a_0$ mass (not considering the disconnected quark diagrams in the calculation). This proceeding is devoted to the further
study on the scalar channel and further discussion on the pseudoscalar. First, we calculate the dispersion relation of the ground state of the scalar to
discern the one-particle state from the possible two-particle state. As the second step, we calculate the correlation function of the isoscalar $q\bar{q}$
operators by taking into account the disconnected diagrams, and then investigate the possible mixing of glueball and $q\bar{q}$ meson by analyzing the correlation
matrix of the gluonic and $q\bar{q}$ operators. As for the flavor singlet pseudoscalar channel, we compare the existing calculations with both full-QCD lattice studies and those from QQCD, and try to address the reason why the correlation functions of the topological charge density operators and
the conventional glueball operators give different ground state mass. The theoretical implication of these different pseudoscalar mass is also discussed.

\section{Lattice setup and numerical details}\label{sec-1}
%\subsection{Gauge configuration details}\label{sec-1-1}
We have generated relatively large gauge ensembles with $N_f=2$ dynamical quarks on anisotropic lattices by the application of USQCD software Chroma~\cite{Edwards:2004sx}. We have two quark masses corresponding to the pion mass around $938$ MeV and $650$ MeV, respectively.
The gauge action we used is the tadpole improved gluonic action~\cite{Morningstar:1997} and anisotropic clover fermion action is utilized in the fermion sector \cite{Edwards:2008ja}.
Table~\ref{conf details} shows the gauge configuration details. In order to determine the lattice spacings $a_s$ at the two pion masses, we calculate the static potential parameterized as $V(r)=V_0+\alpha/r+\sigma r$, and then use the relation $r^2 \frac{dV(r)}{dr}|_{r=r_0}=1.65$ to get $a_s$, where $r_0^{-1}=410(20)$ MeV is the Sommer scale parameter. The advantage of using anisotropic lattice is two-folds: on
the one hand, a large statistics can be obtained by a relatively low cost of
computational resources, on the other hand, the finer lattice spacing in the temporal direction can provide a better resolution for the signal of physical states.
\begin{table}[htb]
	\centering
	\caption{\label{conf details} Parameters of configurations}
	\begin{center}
		\begin{tabular}{cccccc}
			\hline
			\hline
			$\beta$      &      $L^3\times T$    &   $\xi$   &   $a_s$       &   $m_{\pi}$     &    $N_{conf}$\\\hline
			$2.5$        &      $12^3\times128$  &    $5$    &   $0.114fm$   &   $650~\mev$ &    $4800$  \\
			$2.5$        &      $12^3\times128$  &   $5$     &   $0.118fm$   &   $938~\mev$ &    $10400$ \\\hline
			\hline
		\end{tabular}
	\end{center}
\end{table}

As described in~\cite{Sun:2017ipk}, ground state glueball masses of the scalar, tensor, and pseudoscalar are calculated on these two gauge ensembles
by adopting the well-known numerical techniques in calculating the glueball spectrum~\cite{Morningstar:1999rf,Chen:2005mg}. On the lattice, the scalar, tensor, and pseudoscalar correspond to
the lattice quantum numbers $R^{PC}=A_1^{++}$, $(E\oplus T_2)^{++}$, and $A_1^{-+}$, respectively, where $A_1, E, T_2$ are the irreducible representations
of the spatial symmetry group $O$ on the lattice. For each of these quantum numbers, we construct 24 gluonic operators based on different smeared Wilson loops, which give a $24\times 24$ correlation matrix. By solving the generalized eigenvalue problem, we get the optimized gluonic operators which couple mostly to the ground state, and the mass parameter of the ground state can be reliably derived. Ground state masses of these three channels are presented in Table~\ref{compare}, where we also include the previous lattice results for comparison.

\begin{table}[tbp]
	\centering
	\caption{\label{compare} Comparison of ground state scalar, pseudoscalar and tensor glueball mass from quenched, $N_f=2$ and $N_f=2+1$ studies.}
	\begin{center}
		\begin{tabular}{lclll}
			\hline\hline
			&      $m_\pi$ (MeV)    & $m_{0^{++}}$ (MeV)     &   $m_{2^{++}}$ (MeV)     &   $m_{0^{-+}}$ (MeV)      \\
			\hline
			
			$N_f=2$                       &  $938$  &    1397(25)    &    2367(35)             &   2559(50)               \\
			&  $650$  &    1480(52)    &    2380(61)             &   2605(52)               \\
			%%&&&&\\
			$N_f=2+1$~\cite{Gregory:2012hu}&  $360$ &    1795(60)    &    2620(50)             &    ---                    \\
			&&&&\\
			quenched~\cite{Morningstar:1999rf} &        ---          &    1710(50)(80)&    2390(30)(120)        &    2560(35)(120)         \\
			quenched~\cite{Chen:2005mg}        &        ---          &    1730(50)(80)&    2400(25)(120)        &    2590(40)(130)         \\
			
			\hline\hline
		\end{tabular}
	\end{center}
\vspace{-1.5em}
\end{table}
From the table one can see clearly that in the tensor and pseudoscalar channel, both the QQCD studies and the calculations with dynamical quarks give compatible
result of the ground state glueball masses $m_{2^{++}}\sim 2.4-2.6$ GeV, which implies the unquenching effects are not important for the glueball masses.
In contrast, the ground state mass we get in the scalar channel is lower than the result from QQCD and previous $N_f=2+1$ study. In view of this, we perform a further study on the identification of the ground state of the scalar, as described in the following.

\section{The lowest-lying scalar state}\label{sec:3}
\begin{figure}[tp]
	\centering
	\subfigure[$m_\pi\sim938\mev$]%
	{\includegraphics[width=0.4\textwidth,clip]{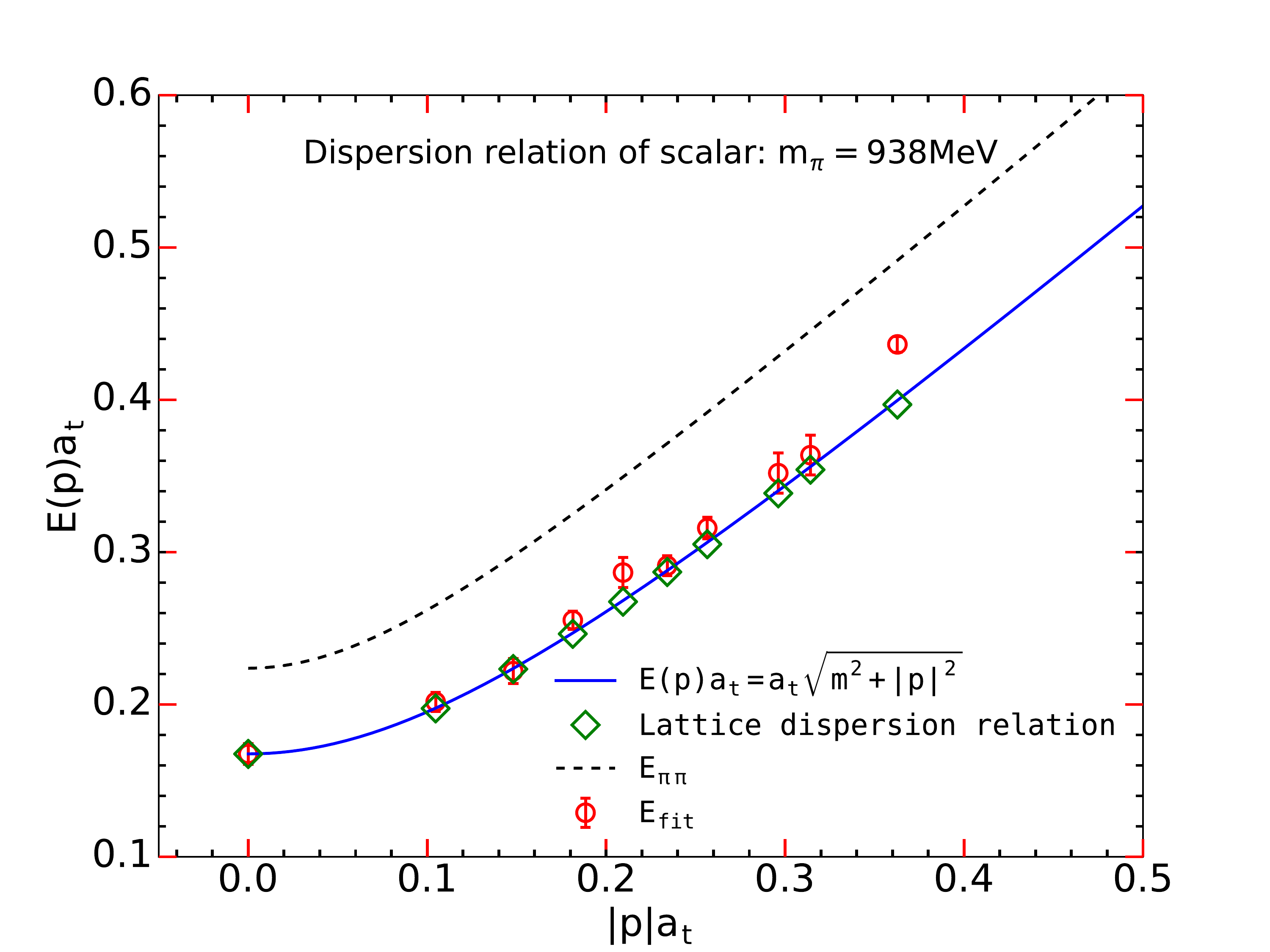}}\hfil
	\subfigure[$m_\pi\sim650\mev$]%
	{\includegraphics[width=0.4\textwidth,clip]{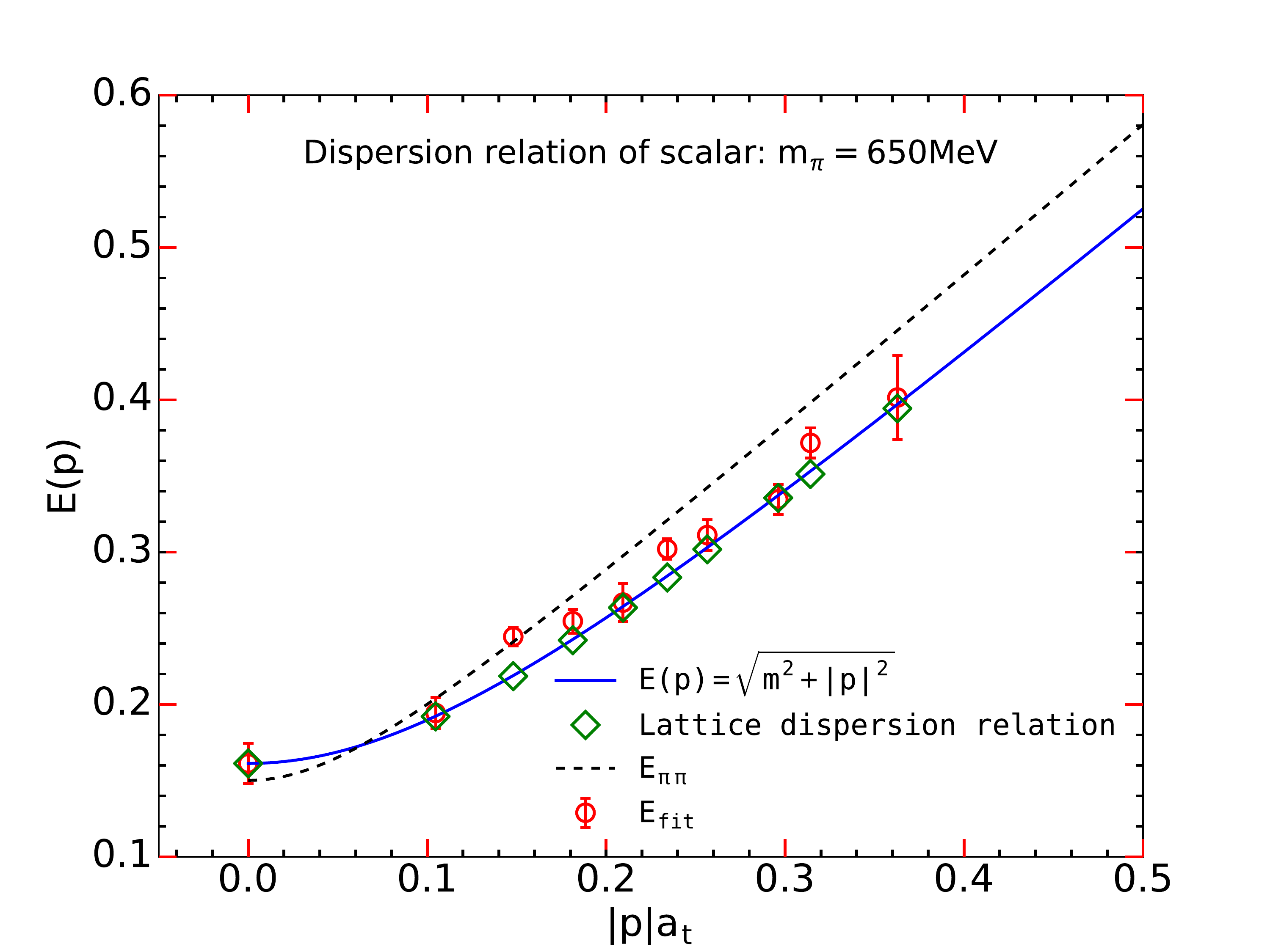}}
	\caption{Dispersion relation of ground scalar states. The red points with error bars are the fitted $E(\mathbf{p})$. The blue solid line and green diamonds points show theoretical continuum and lattice dispersion relation of single particle, respectively, while dash line shows lowest free two pion states with relative momentum $\mathbf{p}$. }
	\label{dispersion}
	\vspace{-1.5em}
\end{figure}

For our gauge configurations with $N_f=2$ dynamical quarks, since there are both quarks and gluons propagating in the sea, the correlation
functions of the scalar operators can have contributions from glueballs, conventional $q\bar{q}$ mesons, and even multi-hadron states, in principle, if
the mixing among them is not strong. If there exists mixing, the eigenstates of the Hamiltonian can be the admixture of these objects. Therefore, in order
to investigate the nature of the ground state, as the first step, we calculate the dispersion relation of the ground state to check whether it is
a single particle state or a multi-hadron state. For the scalar operator set which includes 24 gluonic operators $\{O_\alpha(x),\alpha=1,2,\ldots, 24\}$, we take the optimized
operator
\begin{equation}
O_G^{(1)}(\mathbf{p},t)=\frac{1}{V_3} \sum\limits_{\mathbf{x}} e^{-i\mathbf{p}\cdot \mathbf{x}}\sum\limits_\alpha v_\alpha^{(1)}O_\alpha(x)
\end{equation}
 which couples most to the ground state with momentum $\mathbf{p}$ (here $V_3$ is the spatial volume of the lattice, the combination coefficients $v_\alpha^{(1)}$ correspond to the eigenvector $v$ with the largest eigenvalue $\lambda$ by solving the generalized eigenvalue problem (GEVP) $\mathcal{C}(t)v=\lambda \mathcal{C}(t_0)v$, with $\mathcal{C}(t)$ being the correlation matrix of the operator set). The correlation function for a scalar state with a center-of-mass momentum $\mathbf{p}$ is calculated as
\begin{equation}
C_G(t;\mathbf{p})=\langle O_G^{(1)}(\mathbf{p},t)O_G^{(1)\dagger}(\mathbf{p},0)\rangle\propto e^{-E(\mathbf{p})t}~~~(t\rightarrow \infty),
\end{equation}
where $E(\mathbf{p})$ is the energy of the scalar state with momentum $\mathbf{p}$.  In Figure~\ref{dispersion} we plot the derived $E(\mathbf{p})$ as points
with errorbars. The left panel is for the case of $m_\pi=938$ MeV, and the right one for $m_\pi=650$ MeV. The solid line is the theoretical expectation of the
dispersion relation of a single particle, $E(p)=\sqrt{m^2 + |\mathbf{p}|^2}$,
where $m$ is the ground state mass of the particle, which we obtained on these two gauge configuration ensembles and listed in Table~\ref{compare}. We also show the lattice dispersion relation as diamonds in Figure~\ref{dispersion}, which is calculated similar to the previous formula with the
substitution of $\hat{p}_i=2/a_s \sin (p_i a_s/2)$ for $p_i$. One can see that the two dispersion relations differ little from each other in the momentum region we
are using. For the multi-hadron states, we only consider the lowest $\pi\pi$ states, whose  energies depend on the momentum $\mathbf{p}$ as $E_{\pi\pi}(\mathbf{p})=m_\pi + \sqrt{m_\pi^2+|\mathbf{p}|^2}$ and are shown in the figure by the dashed lines.  For the case of  $m_\pi\sim938$ MeV,
all the data points of $E(\mathbf{p})$  lie on the solid curve and are in good agreement with the single particle dispersion relation. This is exactly what we expect
since $E(\mathbf{p})$ is far below the $\pi\pi$ threshold in this case.  For the $m_\pi\sim650$ MeV case, even though the fitted $E(\mathbf{p})$ are compatible
with both one particle and $\pi\pi$ energy in the small $|\mathbf{p}|$ region, when $|\mathbf{p}|$ increases,  $E(\mathbf{p})$ agrees with the one-particle
dispersion relation better and better and implies that the ground state is (mostly) a single-particle state.

\begin{figure}[tbp]
	\centering
	\sidecaption
	\includegraphics[width=7cm,clip]{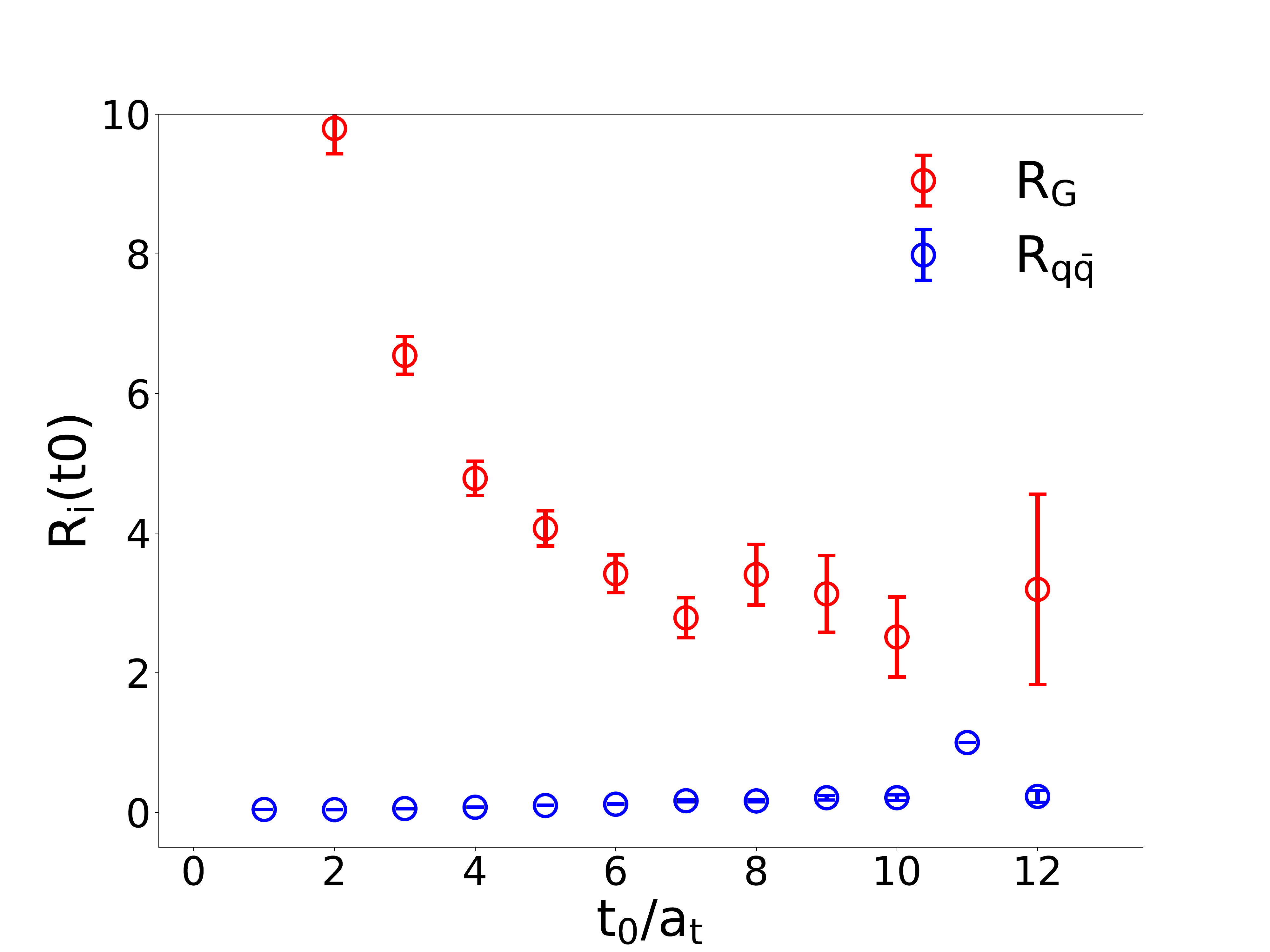}
	\caption{The figure shows the the ratios $R_i(t_0)=\frac{\langle 0|O_i|1^{st}(t_0)\rangle}{\langle 0|O_i|2^{nd}(t_0)\rangle}$ at $m_\pi=938$ MeV. The red points are for the gluonic operator $O_G^{(1)}$ and the blue points for $O_{q\bar{q}}$. $O_G^{(1)}$ couples substantially to both states while the coupling of $O_{q\bar{q}}$ to the first state (likely a glueball state) is relatively very small.}
	\label{eigvec}
	\vspace{-1.em}
\end{figure}

If the scalar ground state can be identified as a single particle state, the second question  is whether it is a glueball, or a $q\bar{q}$ meson, or
an admixture of both.  In order to answer this question, we explore  the possible mixing of the scalar glueball and the scalar $q\bar{q}$ meson
by calculating the $2\times 2$ correlation matrix
\begin{equation}
\mathcal{C}(\mathbf{p},t)=\begin{pmatrix}\langle O_G^{(1)}(\mathbf{p},t) O_G^{(1) \dagger}(\mathbf{p},0)\rangle &
                                                                   \langle O_{G}^{(1)}(\mathbf{p},t)O_{q\bar{q}}^{\dagger}(\mathbf{p},0)\rangle\\
                                      \langle O_{q\bar{q}}(\mathbf{p},t) O_{G}^{(1)\dagger}(\mathbf{p},0)\rangle &
                                      \langle O_{q\bar{q}}(\mathbf{p},t)O_{q\bar{q}}^{\dagger}(\mathbf{p},0)\rangle
\end{pmatrix}\equiv\begin{pmatrix}C_{11}(t) & C_{12}(t)\\ C_{21}(t) & C_{22}(t) \end{pmatrix}
\end{equation}
where $O_{q\bar{q}}$ is the isoscalar quark bilinear operator
\begin{equation}
O_{q\bar{q}}(\mathbf{p},t) =\frac{1}{V_3}\sum\limits_{\mathbf{x}} e^{-i\mathbf{p}\cdot \mathbf{x}} \frac{1}{\sqrt{2}}\left(u\bar{u}(\mathbf{x},t)+d\bar{d}(\mathbf{x},t)\right).
\end{equation}
After Wick contraction, the correlation function $C_{12},C_{21}, C_{22}$ can be expressed explicitly through quark propagator as
\begin{eqnarray}
C_{12}(t)&=& \sqrt{2}\left\langle O_G^{(1)}(\mathbf{p},t) {\rm Tr} \tilde{S}_F(\mathbf{p};0)\right\rangle,~~~
C_{21}(t)=-\sqrt{2}\left\langle {\rm Tr} \tilde{S}_F(\mathbf{p};t)O_G^{(1)}(\mathbf{p},0)\right\rangle \nonumber\\
C_{22}(t)&=&\left\langle 2 {\rm Tr} \tilde{S}_F(\mathbf{p}; t) {\rm Tr} \tilde{S}_F(\mathbf{p}; 0)
         - \frac{1}{V_3}\sum\limits_{\mathbf{x}}e^{-i\mathbf{p}\cdot\mathbf{x}}{\rm Tr} \left[S_F(\mathbf{x},t;\mathbf{0},0)\gamma_5 S_F^\dagger(\mathbf{x},t;\mathbf{0},0)\gamma_5\right]\right\rangle,
\end{eqnarray}
where the trace operation is for the spin and color indices,
$S_F(\mathbf{x},t;\mathbf{y},t')$ is the quark propagator and $\tilde{S}_F(\mathbf{p};t)=\frac{1}{V_3}\sum\limits_{\mathbf{x}}S_F(\mathbf{x},t;\mathbf{x},t)e^{-i\mathbf{p}\cdot\mathbf{x}}$
corresponds to the quark loop diagram (disconnected diagram) at time $t$.  In practice, instead of calculating the all-to-all quark propagator, we compute the wall-source propagator
$S_F^{(w)}(\mathbf{x},t; t')=\sum\limits_{\mathbf{y}}S_F(\mathbf{x},t;\mathbf{y},t')$ at each time $t'$ on each configuration (here we use the QUDA library~\cite{Clark:2009wm} to do the inversion), such that ${\rm Tr} \tilde{S}_F(\mathbf{p};t)$
can be expressed  by
\begin{equation}
{\rm Tr}\tilde{S}_F(\mathbf{p};t)=\frac{1}{V_3}\sum\limits_{\mathbf{x}} e^{-i\mathbf{p}\cdot\mathbf{x}}{\rm Tr} S_F^{(w)}(\mathbf{x},t;t)-(gauge~~ variant~~ terms).
\end{equation}
Note that the gauge variant terms diminish up to $O(1/\sqrt{N_{\rm conf}})$ when averaging over gauge configurations and thus do not contribute to the
correlation function.
\begin{figure}[tbp]
	\centering
	\sidecaption
	\includegraphics[width=7cm,clip]{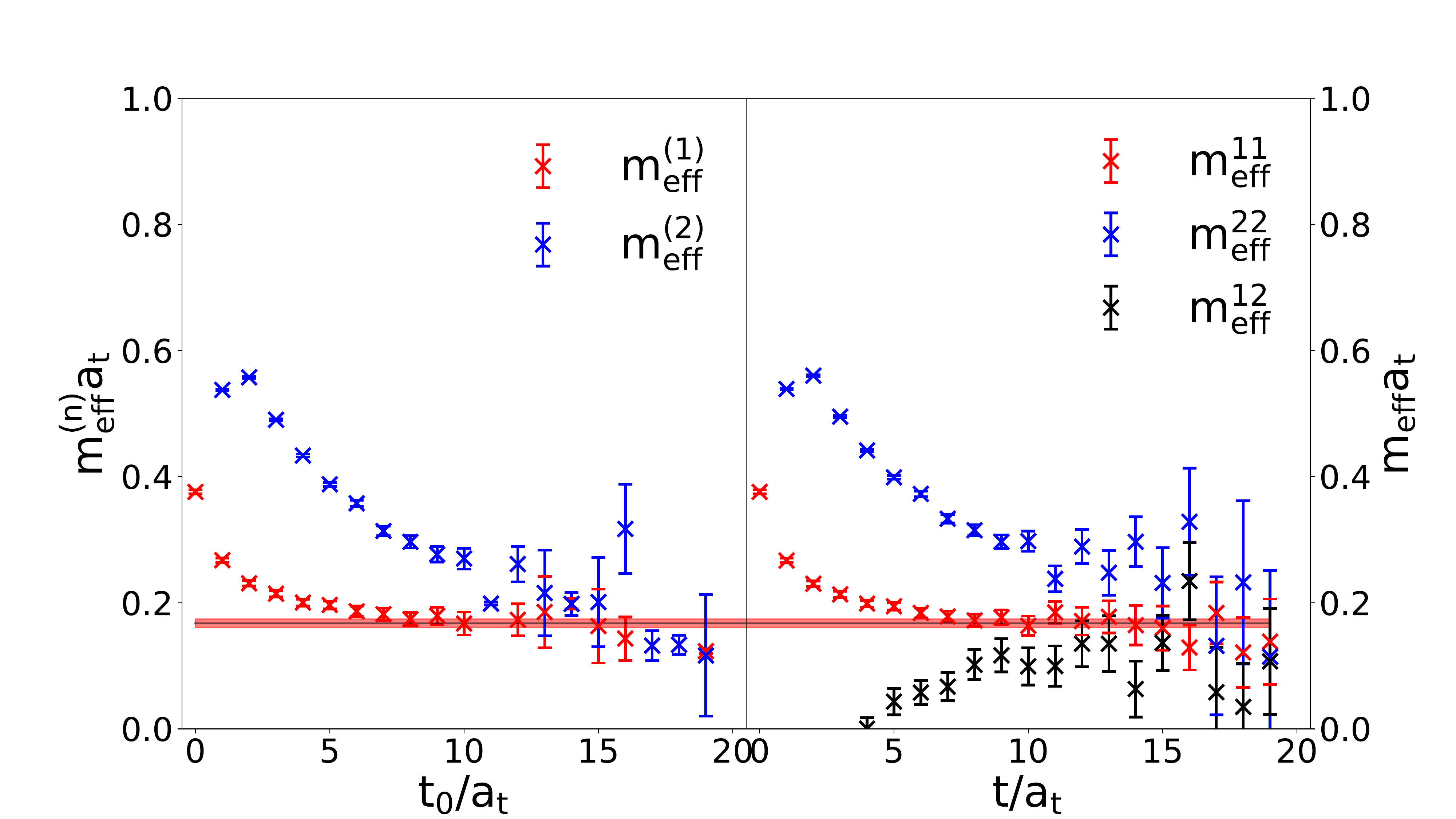}
	\caption{ $m_{\rm eff}^{(n)}(t_0)=-\ln \lambda^{(n)}(t_0)$ are plotted with respect to $t_0$ in the left panel. For comparison, the effective mass of $C_{11}(,t)$,$C_{12}(t)$,$C_{22}(t)$ are plotted in the right panel.
	The error band shows the fitted ground state mass for optimal glueball operators at $m_\pi\sim938$ MeV.}
	\label{eigval}
	\vspace{-1.5em}
\end{figure}%
The effective masses of $C_{11}(t)$,$C_{12}(t)$,$C_{22}(t)$ are plotted in the right panel of Figure~\ref{eigval}. It clearly shows
that $C_{12}$ has significant signals which imply that $O_G^{(1)}$ couples substantially to $O_{q\bar{q}}$. We carry out the GEVP analysis on the $2\times 2$
correlation matrix $\mathcal{C}(\mathbf{0},t)$ (note that $\mathcal{C}$ is not symmetric here because of $C_{21}=-C_{12}$ under the periodic condition in the
time direction)
\begin{equation}
\mathcal{C}(\mathbf{0},t_0+1)v^{(n)}=\lambda^{(n)}(t_0) \mathcal{C}(\mathbf{0},t_0)v^{(n)},
\end{equation}
where $\lambda^{(n)}(t_0)$ and $v^{(n)}(t_0)$ ($n=1,2$) are the $n$-th eigenvalue and eigenvector at $t_0$, respectively. We plot $m_{\rm eff}^{(n)}(t_0)=-\ln \lambda^{(n)}(t_0)$ in the left panel of Figure~\ref{eigval}. In principle, if $\mathcal{C}(\mathbf{0},t)$ had contribution from only two state, $\lambda^{(n)}(t_0)$ and $v^{(n)}(t_0)$ would not depend on $t_0$, but from the figure one can see that $m_{\rm eff}^{(n)}(t_0)$ depends actually on $t_0$. This is not strange because
there are surely quite a lot states contribute. When $t_0$ increases, $m_{\rm eff}^{(2)}$ decreases rapidly and merge into the red band which corresponds to
the ground state mass extracted from the gluonic operators (the effective mass plateau of $C_{11}(t)$). For $m_{\rm eff}^{(2)}(t_0)$ we cannot
infer any meaningful information since there is no clear plateau showing up.  On the other hand, the eigenvector $v^{(n)}(t_0)$ satisfies $v_i^{(n)} v_j^{(n)} C_{ij}(t_0)=\frac{1}{2m_{\rm eff}^{(n)}(t_0) V_3} Z^{(n)}(t_0)Z^{(n)*}(t_0) e^{-m_{\rm eff}^{(n)}(t_0)t_0}$ with $Z^{(n)}(t_0)=\langle 0|v_1^{(n)}O_G^{(1)}+v_2^{(n)}O_{q\bar{q}}|n^{\rm th}(t_0)\rangle$. So we can determine the matrix element $\langle 0|O_i|n^{\rm th}(t_0)\rangle$ by the relation
\begin{equation}
\langle 0|O_i|n^{\rm th}(t_0)\rangle =\frac{v_j^{(n)}C_{ij}(t_0)}{v_i^{(n)} v_j^{(n)} C_{ij}(t_0)}Z^{(n)*}(t_0),
\end{equation}
where $O_i$ refers to $O_G^{(1)}$ and $O_{q\bar{q}}$. We introduce the ratio
\begin{equation}
R_i(t_0)=\frac{\langle 0|O_i|1^{st}(t_0)\rangle}{\langle 0|O_i|2^{nd}(t_0)\rangle}
\end{equation}
to depict difference of the couplings of $O_i$ to the two different states.  Figure~\ref{eigvec} shows these ratios for $O_G^{(1)}$ ( red points) and $O_{q\bar{q}}$(blue points).  When $t_0$ increases, $R_G$ tends to be flat at value of roughly 2.5, which implies that even though it couples largely to the first state, $O_G^{(1)}$ couples substantially to the second state. In contrast, $R_{q\bar{q}}$ is very small (almost zero) all over the $t_0$ region and signals that
$ O_{q\bar{q}}$ coupling almost totally to the second state, but couples little to the first state. This observation gives a strong hint that the first state
might be predominantly a glueball state with a mass of roughly $1.4$ GeV (see in Table~\ref{compare}). This mass seems somewhat lighter than the
scalar glueball mass (roughly 1.7 GeV) from QQCD, but it should be noted that the QQCD result is the extrapolated value at $a_s=0$. It is observed that
the scalar glueball mass has strong dependence on $a_s$~\cite{Morningstar:1999rf,Chen:2005mg}.  In Ref.~\cite{Chen:2005mg}, at $\beta=3.0$ which corresponds to the lattice spacing $a_s=0.119(1)$ fm (very close to the $a_s$ we use in this work) , the scalar glueball mass is roughly $1.5$ GeV and is
not far from the value we mentioned above.

Based on the discussion above, we can draw a preliminary conclusion that the lowest lying scalar state we obtained using the gluonic operator $O_{G}^{(1)}$
is predominantly a glueball state. The $q\bar{q}$ type operator $O_{q\bar{q}}$ couples little to this state. Of course, the pion masses we are using are still
heavy. We will study the quark mass dependence in the future to investigate if the situation changes and how glueball states mix with conventional $q\bar{q}$
states at lighter quark masses.

\section{Flavor singlet pseudoscalar states}

For the calculation of correlation functions of the flavor singlet pseudoscalar, three kinds of lattice operator prototypes can be used, such as
the quark bilinear operator $P(x)=\frac{1}{\sqrt{2}}\left(\bar{u}\gamma_5 u(x)+\bar{d}\gamma_5 d(x)\right)$ (for isospin $SU(2)$) or $P(x)=\frac{1}{\sqrt{3}}\left(\bar{u}\gamma_5 u(x)+\bar{d}\gamma_5 d(x)+\bar{s}\gamma_5 s(x)\right)$  (for flavor $SU(3)$), the topological charge density $q(x)=\frac{1}{32\pi^2}\epsilon_{\mu\nu\rho\sigma}TrF_{\mu\nu}F_{\rho\sigma}(x)$, and the conventional gluonic operators $O_{\rm PS}$ made up of Wilson loops.

Theoretically, the correlation function of $q(x)$ can be expressed as $C_q(x,y)=A\delta(x-y)+\bar{C}_q(|x-y|)$,
where the first term is a contact term which contributes to the positive topological susceptibility, and the second term is negative due to the negative parity of
$q(x)$ and the space-time interchange symmetry in the Euclidean space. Taking $r=|x-y|$, $\bar{C}_q(r)$ can be parameterized as $\bar{C}_q(r)=N\frac{m_{\rm PS}}{4\pi^2r}K_1(m_{\rm PS}r)$~\cite{Shuryak:1995},
which facilitates us to extract the ground state mass $m_{\rm PS}$ from the negative tail of $C_q(r)$.
%Figure~\ref{twopf_eta} shows the $C_q(r)$ we obtained
%on the two gauge ensembles mentioned above (the data points with different colors indicates the different $q(r)$ defined by the smeared gauge links
%at different gradient flow time $t$), which exihibite the expected $r$ behaviors.

We collect the existing lattice results of the mass of the flavor singlet pseudoscalar meson in Table~\ref{pseudo} for an overview.  In the quenched approximation, since there are no sea quark loops, the $P(x)$ type operators are not adopted to extract the mass of the flavor singlet
pseudoscalar mesons. In Ref.~\cite{Chowdhury:2014mra}, the authors use $q(x)$ as pseudoscalar operators and derive the ground state mass $m_{\rm PS}=2.563(34)$ GeV, which is almost the same as the pseudoscalar glueball mass $m_{PS}=2.560(140)$ GeV~\cite{Morningstar:1999rf} and $2.590(140)$ GeV~\cite{Chen:2005mg} in the quenched lattice QCD studies. This is exactly what should be, since there are only pseudoscalar glueball propagating along time
if no valence quarks involved.
%\begin{figure}[tbp]
%	\centering
%	\subfigure[$m_\pi\sim938\mev$]%
%	{\includegraphics[width=0.38\textwidth,clip]{pics/wflow_104.pdf}}\hfil
	%
%	\subfigure[$m_\pi\sim650\mev$]%
%	{\includegraphics[width=0.38\textwidth,clip]{pics/wflow_48.pdf}}
%	\caption{Correlation functions $C_q(r)$ of the topological charge
%		density operators. Different curves correspond to $C_q(r)$ at different Wilson flow time $t=0.2,0.3,0.4$ and $0.8$.}
%	\label{twopf_eta}
%end{figure}
\begin{table}[tbp]
	\centering
	\caption{\label{pseudo} Summary of pseudoscalar states.}
	\begin{center}
		\begin{tabular}{l|ccc}
			\hline
			\hline
			&		$P(x)$	&  $q(x)$	 &  $O_{\rm PS} $ \\\hline
			$N_f=0$	&	 	------	&{\bf 2563(34)} MeV~\cite{Chowdhury:2014mra}	& {\bf 2590(140)} MeV~\cite{Chen:2005mg}\\
			$N_f=2$	&	768(24) MeV~\cite{Helmes:2017ccf}&	890(38) MeV  ~\cite{Sun:2017ipk}	&	{\bf 2605(52)} MeV~\cite{Sun:2017ipk}	\\
			$N_f=2+1$&	947(142) MeV~\cite{Christ:2010dd}	& 1019(119) MeV~\cite{Fukaya:2015ara}	&	------				\\
			$N_f=2+1+1$&	{1006(65)MeV}	~\cite{Michael:2013gka}&	------	   &	------		\\
			\hline\hline
		\end{tabular}
	\end{center}
\end{table}

When dynamical quarks are included in the lattice simulation, the situation is totally different.  There are several works using $P(x)$ to calculate the
$\eta'(\eta_2)$ mass in the lattice simulation with dynamical quarks, and give the results $m_{\eta_2}=768(24)$ MeV ($N_f=2$)~\cite{Helmes:2017ccf},
$m_{\eta'}=947(142)$ MeV ($N_f=2+1$)~\cite{Christ:2010dd}, and $m_{\eta'}=1006(65)$ MeV ($N_f=2+1+1$)~\cite{Michael:2013gka}, respectively, which almost reproduce the experimental
result $m_{\eta'}=958$ MeV. Now when the $q(x)$ operator is applied, $N_f=2+1$ lattice simulation gives the result $m_{\eta'}=1019(119)$ MeV at the physical
pion mass~\cite{Fukaya:2015ara}, which is consistent with the result from $P(x)$ operator. We also calculate the ground state mass using $q(x)$ operator
on our $N_f=2$ gauge configurations and obtain the result $m_{\rm PS}=890(38)$ MeV at $m_\pi=650$ MeV, which is compatible with the $m_{\eta_2}=768(24)$ MeV above (note that our $m_\pi$ is higher than in~\cite{Helmes:2017ccf}). The similar result of $m_{\eta',\eta_2}$ from the operators $P(x)$ and $q(x)$
can be understood as follows. Due to the $U_A(1)$ anomaly, $q(x)$ is now related to $P(x)$ through the
PCAC relation $\partial_\mu A_\mu(x) = 2mP(x) - N_fq(x)$, where $A_\mu(x)$ is the flavor singlet axial vector current, $m$ is the current quark mass. The relation implies that $q(x)$ can couple substantially to
the flavor-singlet $q\bar{q}$ meson, either $\eta_2$ for isospin $SU(2)$ or $\eta'$ for flavor $SU(3)$ (here we assume $\eta'$ as the flavor $SU(3)$
singlet and ignoring the mixing between the isoscalar and the flavor $SU(3)$ singlet).

The interesting thing happens when we use the conventional gluonic operator $O_{\rm PS}$ to calculate the pseudoscalar mass in the dynamical lattice simulation.  It is found that the correlation function of $O_{\rm PS}$ is dominated by the contribution from a much higher state with a mass $m_{\rm PS}=2.559(50)$ GeV at $m_\pi=938$ MeV, and $m_{\rm PS}=2.605(52)$ GeV at $m_\pi=650$ MeV, and there is no clear contribution from $\eta_2$ 
even though it's mass is much lower ($m_{\eta_2}$ is around 1 GeV at the two $m_\pi$'s). Obviously, this heavy pseudoscalar state is very close in mass to
the pseudoscalar glueball in QQCD, and can be taken as the counterpart in the dynamical lattice QCD. In view of this observation, we can tentatively claim that a heavy pseudoscalar glueball (at least in the Hilbert state space) with a mass around 2.6 GeV may exist in the full-QCD, since the existence
of $\eta_2$ manifests the effect of the dynamical quarks.

In order to understand the escape of $\eta_2$ from the correlation functions of $O_{\rm PS}$ type operators, we investigate the explicit form of $O_{\rm PS}$
through the small-$a$ expansion and obtain the expression~\cite{Sun:2017ipk}
\begin{equation}
O_{\rm PS}(x)\propto {\rm Tr}_{\rm color}\epsilon_{ijk}B_i(x) D_j B_k(x)+O(a^2),
\end{equation}
where $B_i(x)$ is the chromo-magnetic field strength operator, $D_i$ is the covariant derivative operator, which is distinct from the continuum form of the
topological charge density $q(x)\propto {\rm Tr}_{\rm color} E_i(x) B_i(x)$ where $E_i(x)$ is the chromo-electric field strength operator. Our observation may
be owing to the fact that $O_{\rm PS}$ couples predominantly to the glueball state, while $q(x)$ can couple to both $\eta'$ ($\eta_2$) state and glueball states.

\section{Summary}
We have investigated the lowest-lying glueball spectrum in $N_f=2$ with two pion mass $m_\pi=650$ MeV and 938 MeV, and mainly focus on the pseudoscalar and scalar channel.
In the flavor singlet scalar channel, we get the lowest state with mass of roughly $1.4$ GeV through gluonic operators, whose dispersion relation is compatible with that of a single particle. We also introduce the $q\bar{q}$ operator to carry out a $2\times 2$ GEVP analysis along with the gluonic operators, and find that the mass of the lowest state does not change. It is also observed that while the gluonic operators couples substantially to both states, the $q\bar{q}$ operator couples  little to the lowest state. This imply that the lowest state might be predominantly a glueball state. The quark mass dependence and the possible mixing of glueball
states with conventional $q\bar{q}$ states will be investigated at smaller pion masses in the future. For the flavor singlet pseudoscalar, we make a joint analysis
on our result and the existing lattice results. It seems that both flavor singlet $q\bar{q}$ meson $\eta'$(or $\eta_2$ for isospin $SU(2)$) and a heavy state with
a mass of roughly $2.6$ GeV exist in the spectrum of lattice QCD with dynamical quarks. The topological charge density operator can couple to both states, which
can be understood by the PCAC relation. In contrast, the conventional gluonic operator for the pseudoscalar glueball couples mainly to the heavy state. Since
its mass is almost the same as the pseudoscalar glueball mass predicted from QQCD lattice study, this heavy state may be a glueball state.

\section*{Acknowledgements}
The numerical calculations are carried out on Tianhe-1A at the National Supercomputer Center (NSCC)
in Tianjin and the GPU cluster at Hunan Normal University. This work is supported in part by the
National Science Foundation of China (NSFC) under Grants No. 11575196, No. 11575197, No. 11335001,
and 11405053. Y.C. and Z.L. also acknowledge the support of NSFC under No. 11261130311 (CRC 110 by
DFG and NSFC). Y. C. thanks the support by the CAS Center for Excellence in Particle Physics
(CCEPP).
%\clearpage
\bibliography{lattice2017}

%%%%%%%%%%%%%%%%%%%%%%%%%%%%%%%%%%%%%%%%%%%%%%%%%%%%%%%%%%%%%%%%%%%%%%%%%%%%%
\end{document}